\documentclass{article}
\usepackage{epsfig}
\usepackage{graphicx}
\usepackage{bm}
\usepackage{amsmath}
\usepackage{cite}

\newcommand{\etal}{\textit{et al.}}
\begin{document}

\title{The Quantum Spin Hall Effect: \\ Theory and Experiment}
\author
{Markus K\"{o}nig$^{1}$, Hartmut Buhmann$^{1}$, Laurens W.
Molenkamp$^{1}$,
\\Taylor L. Hughes$^{2}$, Chao-Xing Liu$^{3,2}$, Xiao-Liang Qi$^{2}$ and Shou-Cheng Zhang$^{2}$\\
\\
\normalsize{$^{1}$Physikalisches Institut (EP III),
Universit$\ddot{\rm a}$t W$\ddot{\rm u}$rzburg}\\
\normalsize{D-97074 W$\ddot{\rm u}$rzburg, Germany}
\\
\normalsize{$^{2}$Department of Physics, McCullough Building,
Stanford
University}\\
\normalsize{Stanford, CA 94305-4045}\\
\normalsize{$^{3}$Center for Advanced Study, Tsinghua University}\\
\normalsize{Beijing, 100084, China}\\
}

\maketitle
\begin{abstract}
The search for topologically non-trivial states of matter has become
an important goal for condensed matter physics. Recently, a new
class of topological insulators has been proposed. These topological
insulators have an insulating gap in the bulk, but have
topologically protected edge states due to the time reversal
symmetry. In two dimensions the helical edge states give rise to the
quantum spin Hall (QSH) effect, in the absence of any external
magnetic field. Here we review a recent theory which predicts that
the QSH state can be realized in HgTe/CdTe semiconductor quantum
wells. By varying the thickness of the quantum well, the band
structure changes from a normal to an ``inverted" type at a critical
thickness $d_c$. We present an analytical solution of the helical
edge states and explicitly demonstrate their topological stability.
We also review the recent experimental observation of the QSH state
in HgTe/(Hg,Cd)Te quantum wells. We review both the fabrication of
the sample and the experimental setup. For thin quantum wells with
well width $d_{QW}< 6.3$~nm, the insulating regime shows the
conventional behavior of vanishingly small conductance at low
temperature. However, for thicker quantum wells ($d_{QW}> 6.3$ nm),
the nominally insulating regime shows a plateau of residual
conductance close to $2e^2/h$. The residual conductance is
independent of the sample width, indicating that it is caused by
edge states. Furthermore, the residual conductance is destroyed by a
small external magnetic field. The quantum phase transition at the
critical thickness, $d_c= 6.3$ nm, is also independently determined
from the occurrence of a magnetic field induced insulator to metal
transition.
\end{abstract}


\section{Introduction}
Our work on the quantum spin Hall (QSH) effect is motivated both by
the quest for spin based electronic devices and by the search for
topologically non-trivial states of matter. Unlike logic devices
based purely on charge current, quantum spintronic devices integrate
information processing and storage units, could operate with low
power consumption and perform reversible quantum
computations~\cite{prinz1998,wolf2001}. Recently, the theoretical
prediction of the intrinsic spin Hall
effect~\cite{murakami2003,sinova2004} has generated great interest
in the field of spintronics, since this effect allows for direct
electric manipulation of the spin degrees of freedom without a
magnetic field, and the resulting spin current can flow without
dissipation. These properties could lead to promising spintronic
devices with low power dissipation. The spin Hall effect has been
observed recently in both $n$~\cite{kato2004} and
$p$~\cite{wunderlich2005} doped semiconductors, however, it is still
unclear if the underlying mechanism is
intrinsic~\cite{murakami2003,sinova2004} or
extrinsic~\cite{dyakonov1971,hirsch1999}.

Beyond the potential technological applications, the intrinsic spin
Hall effect has guided us in the search for new and topologically
non-trivial states of matter. The quantum Hall (QH) state gives the
first, and so far the only example of a topologically non-trivial
state of matter, where the quantization of the Hall conductance is
protected by a topological invariant~\cite{thouless1982}. Given the
fundamental importance of topological quantization in physics, it is
highly desirable to search for quantum states of matter
characterized by non-trivial topological properties similar to, but
distinct from the QH state. Soon after the theoretical prediction of
the intrinsic spin Hall effect in doped semiconductors, a rather
dramatic prediction was made that the intrinsic spin Hall effect
could also be realized in insulators~\cite{murakami2004a}.
Subsequently, the QSH state was independently proposed in
graphene~\cite{kane2005A} and in strained
semiconductors~\cite{bernevig2006A}. The QSH insulator state is
invariant under time reversal, has a charge excitation gap in the 2D
bulk, but has topologically protected gapless edge states that lie
inside the bulk insulating gap. The edge states have a distinct
helical property: two states with opposite spin-polarization
counter-propagate at a given edge~\cite{kane2005A,wu2006,xu2006};
for this reason, they are also called helical edge states. The edge
states come in Kramers' doublets, and time reversal symmetry ensures
the crossing of their energy levels at special points in the
Brillouin zone. Because of this energy level crossing, the spectrum
of a QSH insulator cannot be adiabatically deformed into that of a
topologically trivial insulator without helical edge states;
therefore, in this precise sense, the QSH insulators represent a
topologically distinct new state of matter. The topological
properties of the QSH state are mathematically characterized by a
$Z_2$ topological invariant~\cite{kane2005B}. States with an even
number of Kramers' pairs of edge states at a given edge are
topologically trivial, while those with an odd number are
topologically non-trivial.

While the initial proposal of the QSH state in
graphene~\cite{kane2005A} provided an interesting theoretical toy
model, it was soon shown to be unrealistic since the spin-orbit
gap in this system is extremely small~\cite{yao2007,min2006}.
Bernevig, Hughes and Zhang~\cite{Bernevig2006d} initiated the
search for the QSH state in semiconductors with an ``inverted"
electronic gap, and predicted a quantum phase transition in
HgTe/CdTe quantum wells as a function of the thickness $d_{QW}$ of
the quantum well. The quantum well system is predicted to be a
conventional insulator for $d_{QW}<d_c$, and a QSH insulator for
$d_{QW}>d_c$, with a single pair of helical edge state. In this
paper, we review the basic theory of the QSH state in the
HgTe/CdTe system, and provide explicit and pedagogical discussion
of the helical edge states. We also review recent experimental
observations of the QSH effect in this system, through detailed
discussions of the sample preparation and the experimental setup.

\section{General Properties of (Hg,Cd)Te Quantum Wells}
In this section we will review the bandstructure of bulk HgTe and
CdTe and present a simple model which is useful in describing the
physics of the relevant subbands of a HgTe/CdTe quantum well. HgTe
and CdTe bulk materials have the zinc-blende lattice structure. This
lattice has the same geometry as the diamond lattice, \emph{i.e.},
two interpenetrating FCC lattices shifted along the body diagonal,
but with a different atom on each sublattice. The presence of two
different atoms removes the inversion symmetry of the crystal and
reduces the point group symmetry from $O_h$ (cubic) to $T_d$
(tetrahedral). However, even though the inversion symmetry is
explicitly broken, this only has a small effect on the physics.
Except for the later discussions on the magneto-resistance in the
QSH regime, we will ignore the bulk-inversion asymmetry (BIA).

\begin{figure}[h]
\centering
\includegraphics[scale=0.60]{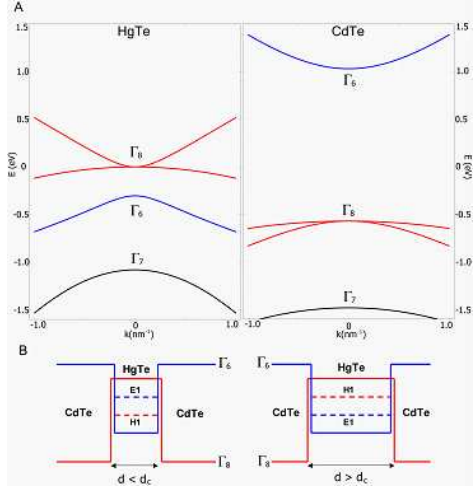}
\caption{(Upper)Bulk bandstructure for HgTe and CdTe (Lower)
Schematic picture of quantum-well geometry and lowest subbands for
two different thicknesses. } \label{bandstructure}
\end{figure}

For both HgTe and CdTe, the important bands near the typical Fermi
level are close to the $\Gamma$-point in the Brillouin zone, and
they are the $s$-type band ($\Gamma_6$) and the $p$-type band which
is split to a $J=3/2$-band ($\Gamma_8$) and a $J=1/2$-band
($\Gamma_7$) by spin-orbit coupling. CdTe, as shown in
Fig.~\ref{bandstructure}~(a), has a band order similar to GaAs with
an $s$-type ($\Gamma_6$) conduction band and $p$-type valence bands
$(\Gamma_8,\Gamma_7)$ which are separated from the conduction band
by a large energy gap of $\sim 1.6$~eV. HgTe as a bulk material can
be regarded as a symmetry-induced semi-metal. Its negative energy
gap of $-300$~meV indicates that the $\Gamma_8$ band, which usually
forms the valence band, is above the $\Gamma_6$ band. The light-hole
bulk subband of the $\Gamma_8$ band becomes the conduction band, the
heavy-hole bulk subband becomes the first valence band, and the
$s$-type band ($\Gamma_6$) is pushed below the Fermi level to lie
between the heavy-hole subband and the spin-orbit split off band
($\Gamma_7$). Based on this unusual sequence of the states, such a
band structure is called inverted. Due to the degeneracy between the
heavy-hole and light-hole bands at the $\Gamma$ point, HgTe is a
zero-gap semiconductor. Note that we have introduced the cubic group
representation labels $\Gamma_6,\Gamma_7,\Gamma_8$ for bands near
the $\Gamma$-point because various orderings of these bands in the
two materials are cumbersome to keep track of and these symbols
explicitly convey the symmetry of the bands. We will use these
symbols for the remainder of the paper instead of referring to
$s$-like, and $p$-like.

\begin{figure}[hbt]
\centering
 \includegraphics[scale=0.85]{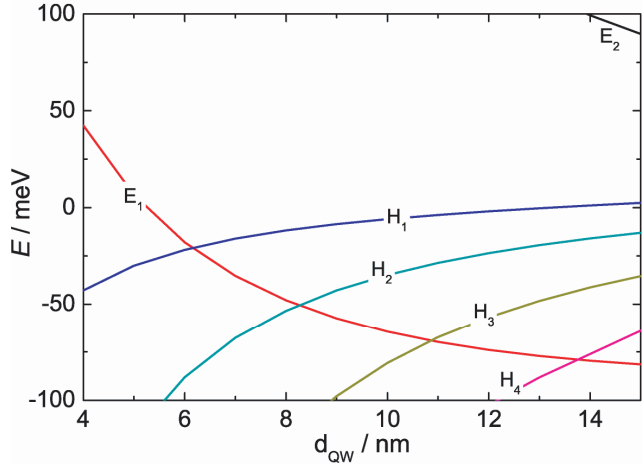}\\
 \caption{The energy of the states in the quantum well are shown as
 a function of the width of the HgTe QW layer.}
 \label{FigEd}
\end{figure}

When HgTe-based quantum well structures are grown, the peculiar
properties of the well material can be utilized to tune the band
structure in a unique way. For wide QW layers, the confinement is
low and the band structure remains ``inverted". However, the
confinement energy increases, when the well width is reduced. Thus,
the energy states will be shifted and, eventually, the energy bands
will be aligned in a ``normal" way, if the QW width falls below a
critical size $d_{c}$. We can understand this heuristically  as
follows: for thin quantum-wells the heterostructure should behave
similarly to CdTe and have a normal band ordering, i.e., the bands
with primarily $\Gamma_6$ symmetry are the conduction subbands and
the $\Gamma_8$ bands contribute to the valence subbands. On the
other hand, as the quantum-well thickness is increased we would
expect the material to look more and more like HgTe which has its
bands inverted. So as the thickness increases we expect to reach a
critical thickness where the $\Gamma_8$ and $\Gamma_6$ subbands
cross and become inverted with the $\Gamma_8$ bands becoming
conduction subbands and $\Gamma_6$ becoming valence subbands. This
is illustrated in Fig.~\ref{bandstructure}~(b). The shift of the
energy levels with QW width can be seen in Fig.~\ref{FigEd}. For the
band structure, self-consistent Hartree calculations have been
performed using a $8\times8$ {\bf k$\cdot$p} model~\cite{Novik05}.
The notation of the subbands as heavy-hole ($H$)-like and
electron ($E$)-like is according to the properties of the respective
wave functions~\cite{Pfeuffer}. The light-hole-like subbands are
energetically remote ($E<-100$~meV) so that they are not depicted in
Fig.~\ref{FigEd}. The transition from a normal band alignment to an
inverted one can clearly be seen in this figure. For a thin QW layer
the quantum confinement gives rise to a normal sequence of the
subbands, i.e., hole-like bands form the valence band and electron
states are in the conduction band. In contrast, a more complex
sequence of the energy states is obtained when the QW width $d_{QW}$
exceeds a critical value $d_{c}\approx 6.3$ nm. In this inverted
regime, the $H_1$ band lies above the $E_1$ subband of the valence
band and, consequently, is now the lowest conduction subband. We
will now develop a simple model and discuss why, immediately after
this crossing, we expect to find a time-reversal invariant bulk
insulating state with topologically protected edge states,
\emph{i.e.}, the QSH state.

\section{Effective Model of the QSHE in HgTe/CdTe Quantum Wells}
The effective model of semiconductors whose outer bands are $s$-like
and $p$-like near the $\Gamma$-point is the $8$-band Kane
model~\cite{kane1957}. This is a model derived from {\bf k$\cdot$p}
perturbation theory for  bulk materials. To calculate the subband
spectrum we must write down a Hamiltonian which is inhomogeneous
along the quantum well growth direction. In
Ref.~\cite{Bernevig2006d} we used this method to derive an effective
$4$-band model for the subbands that participate in the inversion
crossing. Rather than repeat this calculation we will give some
simple symmetry arguments about the form of the model. Under the
cubic symmetry group the relevant $\Gamma_6$-band transforms
antisymmetrically under ``cubic"-parity because it is an
\emph{anti-bonding} band (in CdTe \emph{and} HgTe) , even though it
has an $s$-wave like character. The $\Gamma_8$ band transforms
symmetrically since it is a \emph{bonding} band. The ``conduction"
subbands $E_1,E_2\ldots$ and heavy-hole $H_1,H_2\ldots$ subbands all
alternate their parity under $z\to -z$ inversion.  For our purposes
we neglect the bonding combination of the $\Gamma_6$ and light-hole
bands whose subbands are denoted $L_1,L_2\ldots$ because these bands
are well-separated in energy from the bands in which we are
interested. We will be focussed on only the lowest subbands and as
mentioned before, the $E_1$ subband actually contains components of
the $\Gamma_6$ band and the $\Gamma_8$ light-hole band \emph{i.e.}
$\vert E_1,\pm 1/2\rangle= \alpha(z) \vert \gamma^6,\pm
1/2\rangle+\beta(z)\vert \gamma^8,\pm 1/2\rangle.$ In order to
preserve the transformation of the $E_1$ subbands under inversion of
the growth (z)-direction the envelope function $\alpha(z)$ must be
even under $z\to -z$ while the envelope function $\beta(z)$ piece
must be odd.

The relevant subbands, $E_1$ and $H_1,$ under our assumption of
inversion symmetry,  must be doubly degenerate since time-reversal
symmetry is present. We will label the states with the basis
$\vert E_1 +\rangle,\vert E_1 -\rangle,\vert H_1 +\rangle,\vert
H_1 -\rangle,$ where $\vert E_1 \pm\rangle$ and $\vert H_1
\pm\rangle$ are the two sets of Kramers' partners. The states
$\vert E_1 \pm\rangle$ and $\vert H_1 \pm\rangle$  have opposite
transformations under parity, so a Hamiltonian matrix element that
connects them must be odd under parity since we assumed our
Hamiltonian preserves inversion symmetry. Thus, to lowest order in
$k$, $(\vert E_1 +\rangle,\vert H_1 +\rangle)$ and $(\vert E_1
-\rangle,\vert H_1 -\rangle)$ will each be coupled generically via
a term linear in $k.$ The $\vert H_1 +\rangle)$ heavy-hole state
is formed from the spin-orbit coupled p-orbitals $\vert p_x+ip_y,
\uparrow\rangle)$, while the $\vert H_1 -\rangle)$ heavy-hole
state is formed from the spin-orbit coupled p-orbitals $\vert
-(p_x-ip_y), \downarrow\rangle)$. Therefore, to preserve
rotational symmetry around the growth-axis (z-axis) the matrix
elements must by proportional to $k_{\pm}=k_x\pm ik_y.$ The only
terms allowed in the diagonal elements are terms that have even
powers of $k$ including $k$-independent terms. The subbands must
come in fully degenerate pairs so there can be no matrix elements
between the $+$ state and the $-$ state of the same band. Finally,
we note that if non-zero matrix elements coupled $\vert E_1
+\rangle,\vert H_1 -\rangle$ or $\vert E_1 -\rangle,\vert H_1
+\rangle$ then this would induce a higher-order process that would
couple the $\pm$ states of the same band, thus splitting the
degeneracy. So these matrix elements are forbidden as well. These
simple arguments lead us to the following model
\begin{eqnarray}
{\cal{H}}&=&\left(\begin{array}{cc} h(k)& 0\\
0&h^{*}(-k)\end{array}\right)\label{contH}\\
h(k)&=&\epsilon (k) {\rm{I}}_{2\times 2}+d_a(k)\sigma^a\\
\epsilon (k)&=&C-D(k_{x}^2+k_{y}^2) \\
d_a (k)&=&\left(A k_x,-Ak_y,M(k)\right)\\
M(k)&=& M-B(k_{x}^2+k_{y}^2). \end{eqnarray}\noindent
 where we have used the basis order $({\vert
E_1 +\rangle,\vert H_1 +\rangle,\vert E_1 -\rangle,\vert H_1
-\rangle}),$
 $A,B,C,D,M$ are material parameters that depend on the quantum
well geometry, and we have chosen the zero of energy to be the
valence band edge of HgTe at ${\textbf{k}}=0$ as in
Fig.~\ref{bandstructure}. ${\cal{H}}$ is equivalent to two copies of
the massive Dirac Hamiltonian but with a $k$-dependent mass $M(k)$.

For the purposes of studying the topological properties of this
system, as well as the edge states, we will work with a lattice
regularization of this model which will give us the dispersions over
the entire Brillouin zone torus, i.e., a simplified tight-binding
representation. Since all of the interesting physics at low energy
occurs near the $\Gamma$-point, the behavior of the dispersion at
high-energy is not important. Thus, we can choose a regularization
to make our calculations simple. This simplified lattice model is
\begin{eqnarray}
{\cal{H}}&=&\left(\begin{array}{cc} h(k)& 0\\
0&h^{*}(-k)\end{array}\right)\label{latticeH}\\
h(k)&=&\epsilon (k) {\rm{I}}_{2\times 2}+d_a(k)\sigma^a\nonumber\\
\epsilon (k)&=&C-2Da^{-2}(2-\cos (k_xa)-\cos (k_ya))\nonumber\\
d_a (k)&=&\left(Aa^{-1} \sin (k_xa),-Aa^{-1} \sin (k_ya),M(k)\right)\nonumber\\
M(k)&=&M -2Ba^{-2} \left(2-\cos(k_xa)-\cos (k_ya)\right)\nonumber.
\end{eqnarray}\noindent It is clear that near the
$\Gamma$-point the lattice Hamiltonian reduces to the {\bf
k$\cdot$p} Hamiltonian in Eq.~\ref{contH}. For simplicity, below we
take the lattice constant $a=1$, which corresponds to a redefinition
of the units.

The bulk band structure of this model has the energy spectrum
\begin{eqnarray}
E_{\pm}&=&\epsilon (k)\pm\sqrt{d_a d_a}\\
&=&\epsilon (k)\pm \sqrt{A^2(\sin^2 k_x +  \sin^2 k_y)+M^2 (k)}.
\end{eqnarray} This spectrum gives a bulk insulator as long as
$\sqrt{d_a d_a}\neq 0$ (and no indirect gap is formed) and there is
a gapless spectrum at a phase transition critical point when this
quantity vanishes. The phase transition  and  topological properties
do not depend on $\epsilon (k)$ so we will ignore this term for now.
The lattice Hamiltonian has several critical points, i.e., several
points in $(k_x,k_y,M)$-space where $d_a=0$, but there is only one
that occurs in the low-energy spectrum near the $\Gamma$-point.
Since the model is only valid for small $k$ we will confine
ourselves to this single critical point. We can see from the form of
$d_a$ that this critical point will occur at $k_x = k_y=0$ and $M=0$
(assuming $B$ is finite). For $M$ close to $0$, the gap is minimal
around $\Gamma$ point, and the system is described by two flavors of
Dirac fermions with mass $M$ and $-M$. The Hall conductance of the
massive Dirac model jumps by a quantum ($e^2/h$) when the mass
changes sign~\cite{redlich1984}. That is, $\sigma_H(M\rightarrow
0^+)-\sigma_H(M\rightarrow 0^-)=e^2/h$. For each $2\times 2$ block
of the lattice model (Eq.~\ref{latticeH}), the Hall conductance can
also be calculated explicitly through the Kubo formula, which leads
to $\sigma_H=c_1e^2/h$, with $c_1\in\mathrm{Z}$ the following
Pontryagin winding number~\cite{volovik2003,qi2005}:
\begin{equation}
c_1=-\frac{1}{8\pi^2} \int \frac{d^2
k}{(2\pi)^2}\epsilon_{abc}\epsilon_{ij}\frac{d_a\partial_i
d_b\partial_j d_c}{d^3}
\end{equation}
Direct numerical calculations show that $c_1=0$ for $M/2B<0$ and
$c_1=\pm 1$ for $0<M/2B<2$, where $+(-)$ corresponds to the $h(k)$
($h^*(-k)$) block. Consequently, the system with $M/2B>0$ is
equivalent to two independent quantum Hall systems with opposite
Hall conductance, which guarantees a pair of counter propagating
edge states on each boundary of the system. (The condition $M<4B$
is always satisfied for the physical system since the gap $2M$ is
always small.) Once such a pair of edge states is established on
the boundary they are robust under time-reversal invariant
perturbations, and are protected by time-reversal symmetry which
generates a $Z_2$ topological quantum number. In other words, the
system (Eq.~\ref{latticeH}) with $M/2B>0$ is a topologically
nontrivial insulator which cannot be adiabatically tuned to a
trivial insulator. More discussions about the topological
stability of this system will be presented in the next section.

So far we have developed a simple $4$-band lattice model for the
$E_1$ and $H_1$ subbands and studied the properties of the bulk band
structure and the low-energy critical point.  We will now solve the
model in a finite strip geometry to explicitly show the existence of
topologically protected edge states for the non-trivial regime.
Since the identity term $\epsilon(k)$ in the Hamiltonian
(Eq.~\ref{latticeH}) provides only an overall energy shift to all
the states without changing the wavefunction profiles, we will omit
this term below. The Hamiltonian (Eq.~\ref{latticeH}) can be
rewritten as
\begin{equation}
{\cal{H}}=\sum_{\textbf{k}}\left( A \sin (k_x)\Gamma^1 + A \sin
(k_y)\Gamma^2+{\cal{M}}(\textbf{k})\Gamma^5\right)c^{\dagger}_{\textbf{k}}c_{\textbf{k}}\end{equation}\noindent
where $\Gamma^a,\; a=1\ldots 5$ form a Clifford algebra and
${\cal{M}}(\textbf{k})$ is defined in Eq.~\ref{latticeH}. For this
model we have $\Gamma^1=\sigma^x\otimes
s^z,\;\Gamma^2=-\sigma^y\otimes 1,\;\Gamma^3=\sigma^x\otimes
s^x,\Gamma^4=\sigma^x\otimes s^y,\;\Gamma^5=\sigma^z\otimes 1$ where
$\sigma^i$ acts on the $(E_1,H_1)$ space and $s^i$ acts on $(+,-)$
space. As stated above, around the $\Gamma$-point in the Brillouin
zone this Hamiltonian behaves as a massive Dirac Hamiltonian in
$2$-d which means that for non-zero $M$ the bulk spectrum will be
gapped. Now we want to show that for certain values of the parameter
$M$ the system is in a topological insulator state characterized by
a bulk energy gap and edge-states which lie in the gap. In order to
study the edge-state spectrum we must first pick a particular edge.
Since this simplified model has square lattice symmetry we pick the
edges to be perpendicular to the $y$-axis. This implies that $k_y$
is no longer a good quantum number since the system is not
homogeneous in this direction, but instead has boundaries on the
lines $y=0,L.$ We must Fourier transform $k_y$ into position space
to obtain a $1$-d tight-binding model which depends on the other
momentum $k_x.$ We make the substitution\begin{equation}
c_{\textbf{k}}=\frac{1}{L}\sum_{j}e^{i k_y j}c_{k,j}
\end{equation}\noindent where $k\equiv k_x.$ This
gives the Hamiltonian:
\begin{eqnarray}
{\cal{H}}&=&\sum_{k,j}\left({\cal{M}}c^{\dagger}_{k,j}c_{k,j}+{\cal{T}}c^{\dagger}_{k,j}c_{k,j+1}+
{\cal{T}}^{\dagger}
c^{\dagger}_{k,j+1}c_{k,j}\right)\\
{\cal{M}}&=&A \sin (k_x)\Gamma^1 -2B[2-M/2B-\cos
(k_x)]\Gamma^5\\
{\cal{T}}&=&\frac{i A}{2}\Gamma^2+B\Gamma^5
\end{eqnarray}

We expect edge states to be exponentially localized on the boundary
so we look for solutions with the following
ansatz~\cite{creutz1994,creutz2001}
\begin{equation}
\psi_{\alpha}(j)=\lambda^j \phi_{\alpha}\label{ansatz}
\end{equation}\noindent where $\lambda$ is a complex number, $j$ is
the index of the $j$-th lattice site in the $y$-direction which runs
from $0\ldots L$,  and $\phi_{\alpha}$ is a constant $4$-component
spinor with $\alpha=1\ldots 4.$ Thus $H\psi_{\alpha}=E
\psi_{\alpha}$ gives
\begin{equation}
\lambda^{-1}{\cal{T}}_{\alpha\beta}\phi_{\beta}+\lambda{\cal{T}}^{\dagger}_{\alpha\beta}\phi_{\beta}+{\cal{M}}_{\alpha\beta}\phi_{\beta}=E\phi_{\beta}.
\end{equation}\noindent This Hamiltonian is particle-hole symmetric,
 and time-reversal symmetric. Thus, at
non-zero $M$ a gap is opened around $E=0$ and at $k_x=0$ we expect
the edge states to have energy $E=0.$ So, at the $\Gamma$-point we
generate the following simple equation
\begin{equation}
\left[\frac{i
A}{2}\left(\lambda^{-1}-\lambda\right)\Gamma^2+B\left(\lambda+\lambda^{-1}\right)\Gamma^5+{\cal{M}}(0)\Gamma^5\right]\phi=0.\end{equation}
\noindent Multiplying both sides by $\Gamma^5$  gives
\begin{equation}
\frac{A}{2}\left(\lambda^{-1}-\lambda\right)\left(i\Gamma^5\Gamma^2\right)\phi=-\left(B(\lambda+\lambda^{-1})+{\cal{M}}(0)\right)\phi
\label{evalue}\end{equation}\noindent The operator
$i\Gamma^5\Gamma^2$ has eigenvalues $\pm 1$. First consider
$i\Gamma^5\Gamma^2\phi=-\phi$, under which condition
Eq.~\ref{evalue} becomes a quadratic equation for $\lambda$ which is
easily solved to give
\begin{equation}
\lambda_{(1,2)}=\frac{-{\cal{M}}(0)\pm\sqrt{{\cal{M}}^2(0)+(A^2-4B^2)}}{A+2B}.
\end{equation}\noindent It is trivial to see that if $\lambda$ is
a solution for $i\Gamma^5\Gamma^2\phi=-\phi$ then $\lambda^{-1}$
is a solution for the opposite eigenvalue
$i\Gamma^5\Gamma^2\phi=\phi$. Denoting $\phi_{s+(-)}$, with
$s=1,2$ being the two eigenstates of $i\Gamma^5\Gamma^2$ with
eigenvalue $+1(-1)$, then a generic solution of the Schroedinger
equation can be written as
\begin{eqnarray}
\psi(j)=\sum_s\left(a_s\lambda_{(1)}^j+b_s\lambda_{(2)}^j\right)\phi_{s+}+\sum_s\left(c_s\lambda_{(1)}^{-j}+d_s\lambda_{(2)}^{-j}\right)\phi_{s-}
\end{eqnarray}
The open boundary condition can be expressed as
$\psi_\alpha(j=0)=0$. Since $\phi_{s\pm}$ are mutually orthogonal,
this condition leads to $a_s+b_s=0,~c_s+d_s=0$. On the other hand,
the requirement of normalizability requires that only the
exponentially decaying terms be included in $\psi(j)$.
Consequently, a normalizable edge state solution exists only when
$|\lambda_{(1)}|<1,~|\lambda_{(2)}|<1$ (with $c_s=d_s=0$) or
$|\lambda_{(1)}|>1,~|\lambda_{(2)}|>1$ (with $a_s=b_s=0$).

Such a derivation can be easily generalized to $k_x\neq 0$ case.
Since $\left[i\Gamma^5\Gamma^2,\Gamma^1\right]=0$ the term
$A\sin(k_x)\Gamma^1$ can be simultaneously diagonalized to add to
the energy. The term involving $\cos(k_x)$ simply contributes to the
mass term. It is convenient to choose the eigenstates $\phi_{s\pm}$
so that $\Gamma^1\phi_{s\pm}=s\phi_{s\pm}$, in which condition the
two edge state wavefunctions $\psi_s(j)$ and corresponding energy
$E_s(k_x)$ are written as
\begin{eqnarray}
\psi_s(j)&=&\sum_s\left(\lambda_{(1)}^j-\lambda_{(2)}^j\right)\phi_{s+}\\
E_s(k_x)&=&
-As\sin(k_x)\label{edgedispersion}\\
\text{with~}\lambda_{(1,2)}&=&\frac{-m(k,M)\pm\sqrt{m(k,M)^2+(A^2-4B^2)}}{A+2B}\nonumber\\
m(k,M)&=&-2B(2-M/2B-\cos (k_x))\nonumber. \end{eqnarray} To have a
normalizable solution we still must have $|\lambda^{(1,2)}|<1$ or
$|\lambda^{(1,2)}|>1$, which leads to
\begin{equation}
-2B<m(k,M)<2B\label{ineqcond}
\end{equation}\noindent In other words, for a fixed $M$ an edge state will
exist only for the finite region of $k$ that satisfies this
inequality.

\begin{figure}[h]
\centering
\includegraphics[scale=0.50]{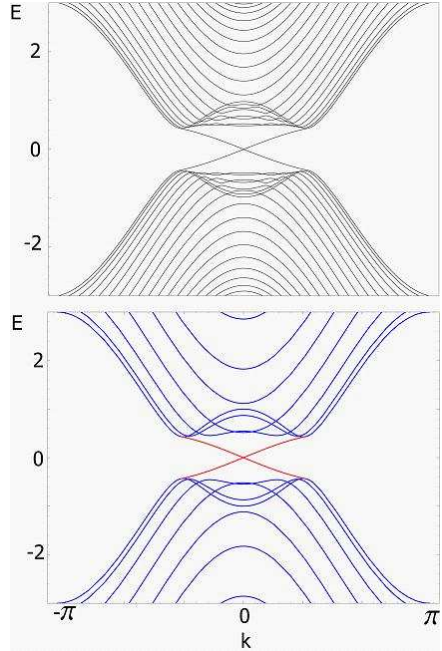}
\caption{(Upper)Exact diagonalization of $1$-d tight binding model.
(Lower) analytic solution for bulk (blue) and edge states(red). Bulk
bands are plotted vs $k_x$ for several different values of $k_y$ in
order to reproduce the look of the exact diagonalization results. }
\label{edgecomb}
\end{figure}

For the dispersion in Eq.~\ref{edgedispersion} we can see that two
edge states with opposite $\Gamma^1$ eigenvalue are propagating in
opposite directions. At $k_x=0$, the edge states exist for
$0<M<4B$, that is, $m(0,M)$ satisfies the inequality in
Eq.~\ref{ineqcond} for $M/2B\in (0,2)$. When $k_x$ is varied from
$0$ to $\pi$, the edge states are dispersive until the inequality
(Eq.~\ref{ineqcond}) is violated at some finite $k_x$, which,
physically, is the wavevector where the edge states touch the bulk
(extended) states. In Fig.~\ref{edgecomb} we have plotted the
energy spectrum as a function of $k_x$ from (numerical) exact
diagonalization, and from our analytic solution. In our analytic
solution, and with fixed $M$, we solve for the $k_x$-points where
the edge state solution is no longer normalizable, which occurs
when the edge state becomes degenerate with the bulk states. It is
at these $k$-points where the edge state dispersion enters the
bulk and is no longer visible.

\section{Topological Properties of the Helical Liquid}

Similar to the case of quantum Hall effect, the topological
properties of the QSH system can be understood by studying the low
energy effective theory of the edge states. The edge states of the
quantum Hall effect are described by a chiral Luttinger liquid
theory~\cite{wen1990}. Such a chiral theory only contains, say,
left movers, so that the current carried by the edge excitations
cannot be destroyed by any perturbation due to the absence of
counter-propagating modes in which to backscatter. In comparison,
the effective theory of the quantum spin Hall edge states do
contain both left and right movers, and can be written as
\begin{eqnarray}
H=\int
\frac{dk}{2\pi}\left(\psi^\dagger_{k+}vk\psi_{k+}-\psi^\dagger_{k-}vk\psi_{k-}\right)+H_{\rm
pert}\label{Effedge}
\end{eqnarray}
where $H_{\rm pert}$ represents possible perturbation terms,
including disorder and/or electron-electron interactions.

If time-reversal symmetry is not present, a simple ``mass" term
can be added in $H_{\rm pert}$ so that the spectrum becomes
gapped:
\begin{eqnarray}
H_{\rm
mass}=\int\frac{dk}{2\pi}m\left(\psi^\dagger_{k+}\psi_{k-}+h.c.\right)\nonumber
\end{eqnarray}
However, the time-reversal symmetry of the electron system is
expressed as
\begin{eqnarray}
T^{-1}\psi_{k+}T=\psi_{-k,-},~T^{-1}\psi_{k-}T=-\psi_{-k,+}
\end{eqnarray}
which implies
\begin{eqnarray}
T^{-1}H_{\rm mass}T=-H_{\rm mass}.\nonumber
\end{eqnarray}
Consequently, $H_{\rm mass}$ is a time-reversal symmetry breaking
perturbation. More generally, if we define the ``chirality" operator
\begin{eqnarray}
C=N_+-N_-=\int
\frac{dk}{2\pi}\left(\psi_{k+}^\dagger\psi_{k+}-\psi_{k-}^\dagger
\psi_{k-}\right)\nonumber,
\end{eqnarray}
any operator that changes $C$ by $2(2n-1),~n\in\mathrm{Z}$ is
time-reversal odd. In other words, time-reversal symmetry requires
that the Hamiltonian $H_{\rm pert}$ only include  processes of
$2n$ particles back scattering, such as
$\psi_{k+}^\dagger\psi_{k'+}^\dagger\psi_{p-}\psi_{p'-}$.
Therefore, the most relevant perturbation in normal metals, i.e.,
$\psi_{k+}^\dagger\psi_{k'-}$, is forbidden by time-reversal
symmetry. This is essential for the topological stability of the
edge states. This edge state effective theory is thus non-chiral
and is qualitatively different from the usual spinless or spinful
Luttinger liquid theories. It can be considered as aa example of a
new class of one-dimensional critical theories, describing a
``helical liquid"~\cite{wu2006,xu2006}. Specifically, for the
non-interacting case no time-reversal invariant perturbation is
available to induce back scattering, so that the edge state is
robust under any disorder.

\begin{figure}[h!]
\centering
\includegraphics[scale=0.42]{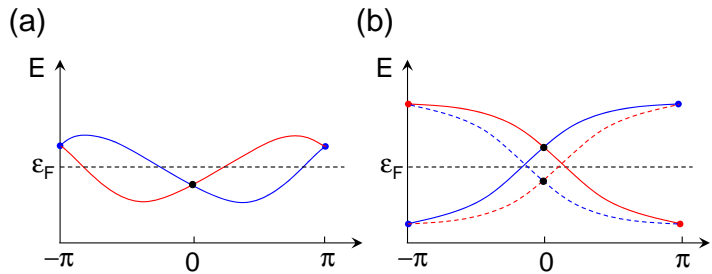}
\caption{(a) The energy dispersion of a one-dimensional
time-reversal invariant system. Kramer's degeneracy is required at
$k=0$ and $k=\pi$, so that the energy spectrum always crosses the
Fermi level $\epsilon_F$  $4n$ times. (b) The energy dispersion of
the helical edge states on one boundary of the QSH system (solid
lines). At $k=0$ the edge states are Kramers' partners of each
other, while at $k=\pi$ they merge to the bulk and pair with the
edge states of the other boundary (dash lines). In both (a) and (b),
red and blue lines form Kramers' partners of each other.}
\label{nogotheorem}
\end{figure}

\begin{figure}[h]
\centering
\includegraphics[scale=0.45]{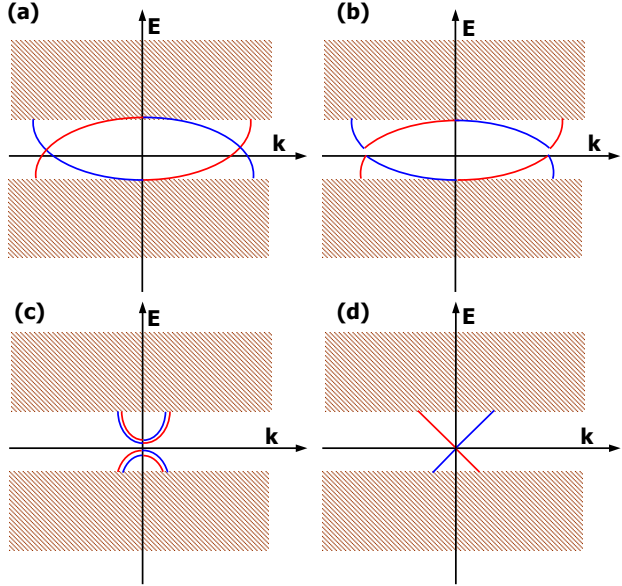}
\caption{Brown areas are regions of bulk states. Red and blue
dispersing edge states represent Kramers' partners (a)An even number
of pairs of fermion branches crossing at $k$ points which do not
transform to themselves under time-reversal (b) A slight
perturbation added to the system causes the edge states in (a) to
form a gap. Kramers' theorem is still satisfied but these edge
states are unstable to gap formation (c) Two pairs of fermion
branches that initially crossed at the special point
${\textbf{k}}=0$ are shown after an infinitesimal perturbation is
added. A gap is formed and Kramers' theorem is satisfied but this
configuration is also unstable to gap formation (d)A single pair of
fermion branches crosses at ${\textbf{k}}=0$. A perturbation cannot
open a gap because in that case there would be two states which were
singly degenerate which will not satisfy Kramers' theorem, thus this
configuration is stable. } \label{edgegapschem}
\end{figure}

Now consider the case with two flavors of helical edge states on
the boundary, {\em i.e.}, a 1-d theory with two left movers and
two right movers, with the Hamiltonian
\begin{eqnarray}
H=\int\frac{dk}{2\pi}\sum_{s=1,2}\left(\psi^\dagger_{ks+}v_sk\psi_{ks+}-\psi^\dagger_{ks-}v_sk\psi_{ks-}\right).\nonumber
\end{eqnarray}
Due to the same reason as the one-flavor case, the single particle
back scattering terms between Kramers' doublets
$\psi^\dagger_{ks+}\psi^{\ }_{-k,s-}$ is forbidden. However, the
back scattering between different flavors such as
$\psi^\dagger_{k1+}\psi^{\ }_{-k,2-}$ can be non-vanishing, which
makes such a theory qualitatively equivalent from a spinful
Luttinger liquid with four degrees of freedom. A mass term like
$\int \frac{dk}{2\pi}M\left(\psi^\dagger_{k1+}\psi^{\
}_{k2-}+h.c.\right)$ can immediately open a gap in the system,
which implies that the existence of gapless edge states is not
generic, as shown in Fig. \ref{edgegapschem}. In other words, two
copies of the helical liquid becomes a topologically trivial
theory. More generally, an edge system with time-reversal symmetry
is a nontrivial helical liquid when there is an odd number of
left(or right) movers, and trivial when there is an even number.
Such a property makes it natural to characterize the topology in
the QSH systems by a $Z_2$ topological quantum number.

There is also an alternative way to understand the qualitative
difference between even and odd branches, which is expressed as a
``no-go" theorem in Ref.~\citen{wu2006}: There is always an even
number of Kramers' pairs at the Fermi energy for an arbitrary
one-dimensional band structure. Such a no-go theorem is a
time-reversal invariant analog of the Nielsen-Ninomiya no-go
theorem for chiral fermions on a
lattice~\cite{NIELSEN1981,NIELSEN1981A}. For spinless fermions
there are always an equal number of left movers and right movers
at the Fermi level, which leads to the fermion doubling problem in
odd spatial dimensions. Similarly, for a time-reversal symmetric
system with half-odd integer spin, Kramers' theorem requires each
energy band to be accompanied by its Kramers' partner, so that the
number of low energy channels are doubled again. As shown in Fig.
\ref{nogotheorem} (a), a Kramers' pair of states at $k=0$ must
re-form into pairs with each other when $k$ goes from $0$ to $\pi$
and $2\pi$, which requires the bands to cross the Fermi level $4n$
times. Seemingly disheartening, there is actually an exception to
this theorem analogous to the reason why a chiral liquid exists in
the quantum Hall effect. The exception is that a helical liquid
with an odd number of fermion branches \emph{can} occur if it is
holographic, i.e., appears at the boundary (edge) of a $2$d bulk
system. As shown in Fig. \ref{edgecomb}, in this case the edge
states are Kramers' partners of each other at $k=0$, but they
merge into bulk states at some finite $k_c$, so that they don't
have to be paired with each other at $k=\pi$. More accurately, at
$k>k_c$ the edge states on both the left and right boundaries
become bulk states at $k>k_c$, and form a Kramers' pair with each
other, as shown in Fig. \ref{nogotheorem} (b).

The no-go theorem also provides a physical understanding of the
topological stability of the helical liquid. Any local
perturbation on the boundary of a 2d QSH system can be considered
as coupling a ``dirty surface layer" with the unperturbed helical
edge states. Whatever perturbation is considered, the ``dirty
surface layer" is always a one-dimensional system, so that there
are always an even number of Kramers' pairs of low energy
channels. Since the helical liquid has only an odd number of
Kramers' pairs, the coupling between them can only annihilate an
{\em even} number of Kramers' pairs, so that at least one-pair of
gapless edge states can survive.

Similar to the correspondence of chiral edge states and bulk Chern
number in quantum Hall effect~\cite{thouless1982}, the $Z_2$
topological stability of helical edge states can be determined by a
$Z_2$ topological quantum number defined in the two-dimensional
Brillouin zone, as shown in Refs.~\citen{kane2005B} and
\citen{fu2006}. Several alternative ways to define the $Z_2$
invariant and its generalization to three-dimensional topological
insulators were also proposed in the
literature~\cite{fu2007a,fu2007b,moore2007,roy2006a,roy2006b,roy2006c}.
In the quantum Hall case, the Chern number can be defined by
introducing twisted boundary conditions, even when disorder and
interactions are present so that momentum is not good quantum
number~\cite{niu1985}. Similar twisted boundary conditions can be
introduced to define the $Z_2$ invariant when disorder is
considered~\cite{essin2007}. However, the generalization to the
many-body case with interactions is not so straightforward, since
the Kramers' degeneracy is only defined for a single electron state.
Some efforts have been made toward this direction
recently~\cite{lee2007}.

In summary, in this section we have discussed  the topological
stability of the edge states, which is essential for the quantum
spin Hall effect in HgTe quantum wells to be realized in an
experimental system. Although the edge states are non-chiral, one
can still detect them through the longitudinal conductance it
contributes. In the ballistic limit, a  helical liquid provides a
metallic one-dimensional channel, which leads to a conductance of
$e^2/h$. For a system with two boundaries the net longitudinal
conductance will be $2e^2/h$. In the rest of the paper we will
review the experimental observations of the transport properties
of HgTe quantum wells and present  theoretical analysis of the
results.

\section{ Quantum Well Fabrication}

\begin{figure}[htb]
\centering
 \includegraphics[width=0.8\linewidth]{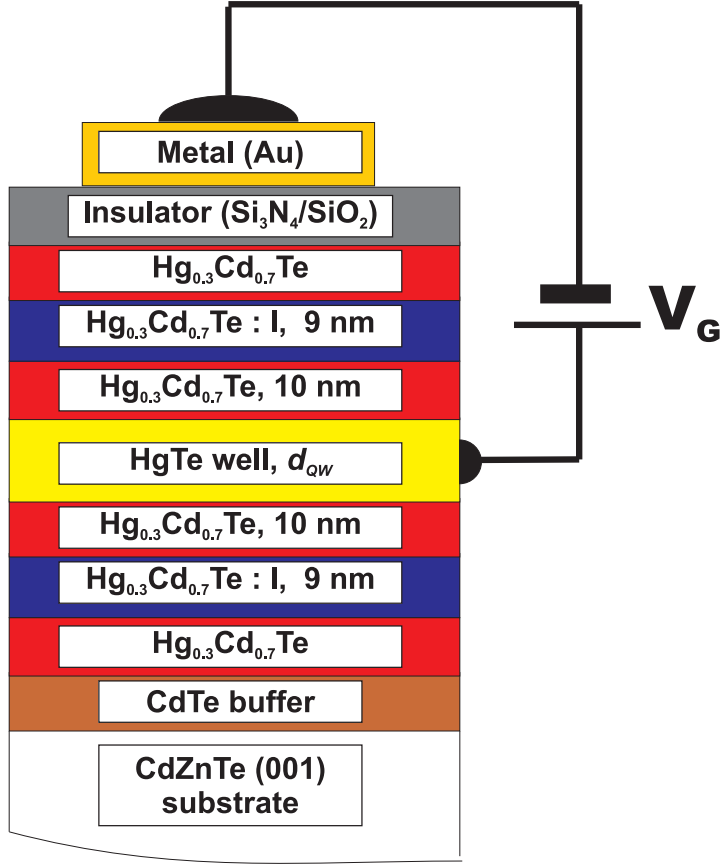}\\
 \caption{Schematics of the layer sequence of the MBE-grown quantum well
 structures.}
 \label{FigMBE}
\end{figure}

As discussed above, we anticipate that the transition from a normal
to an inverted band structure in HgTe QWs coincides with a phase
transition from a trivial insulator to the QSH insulator.
Experimentally, we have investigated the transport properties of
HgTe samples  with QW width in the range from 4.5 nm to 12.0 nm, so
as to cover both the normal and the inverted band structure regime.
The samples were grown by molecular beam epitaxy (MBE) and the layer
sequence is schematically depicted in Fig.~\ref{FigMBE}. The wells
are sandwiched by Hg$_{0.3}$Cd$_{0.7}$Te barriers and  $n$-type
modulation doped using I-doping on both sides of the QW layer.
Recent advances in the growth of HgTe-based QW structures, that were incorporated in these
structures are documented in Ref.~\cite{Becker07}. For example, increasing
the spacer width between the quantum well and the doping layer
results in an enhancement of the carrier mobility $\mu$, and samples
with mobilities of several $10^5~{\rm cm}^2/{\rm (Vs)}$ even at low
densities $n<5\times 10^{11}~{\rm cm}^{-2}$ have been used for the
actual measurements. In such samples, the elastic mean free path is of the
order of several microns.

The devices have been structured by means of optical and electron
beam (e-beam) lithography. So as to avoid Hg evaporation, the HgTe
layer has to be grow at temperatures as low as 180$^{\circ}$C and
the sample has to be kept well below this temperature during the
entire process of fabrication. Standard processes based on PMMA
(polymethylmethacrylate) as a resist can not be used, since these
usually demand temperatures up to 200$^{\circ}$C for bake-out. While
we fabricated our first HgTe-based nanostructures using a standard
photoresist for the e-beam steps~\cite{Daumer03}, we have, for the
samples discussed here, employed a dedicated low temperature PMMA
technology. In combination with Ar-ion etching we can reproducibly
fabricate structures with dimensions down to ca.\ 100~nm.

\begin{figure}[hbt]
\centering
 \includegraphics[width=0.8\linewidth]{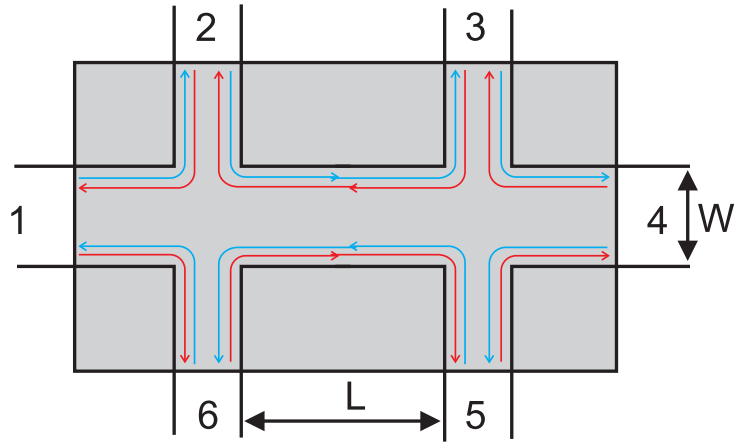}\\
 \caption{This schematic sketch of the devices shows the QSH edge
 states under the gate (shaded region). The ungated parts of the
 sample are $n$-type.}\label{FigHallbar}
\end{figure}

For the investigation of the QSH effect, devices in a Hall bar
geometry of various dimensions have been fabricated from QW structures
with a QW width of 4.5 nm, 5.5 nm, 6.4 nm, 6.5 nm, 7.2 nm, 7.3 nm,
8.0 nm and 12.0 nm, respectively. The layout of the devices is
sketched in Fig.~\ref{FigHallbar}.

A 110~nm thick layer consisting of a Si$_3$N$_4$/SiO$_2$
superlattice is deposited on top of the structure by plasma-enhanced
chemical vapor deposition (PECVD). This layer serves as a gate
insulator between the semiconductor and the (Ti/Au) gate electrode
on top of the structure. The top gate covers the leads of the Hall
bar only to a small extent, while the largest parts of the leads are
not gated, which implies that the mesa is always contacted by
$n$-type HgTe leads. The gate insulator has excellent break-through
properties and allows gate voltages in the range from -5~V to +5~V
which is sufficient to induce a large change in carrier
concentration ~\cite{Zhang01,Becker07}. However, in most structures
not the full voltage swing can be applied. Hinz~\etal\ have reported
hysteresis effects in HgTe QWs at extreme gate
voltages~\cite{Hinz06}. These effects are attributed to the filling
and emptying of trap levels between the insulator and the
semiconductor. In the experiments described below, care was taken to
keep the gate voltages in the non-hysteretic regime.

\section{High Field Characterization}

\begin{figure}[htb]
\centering
 \includegraphics[width=0.8\linewidth]{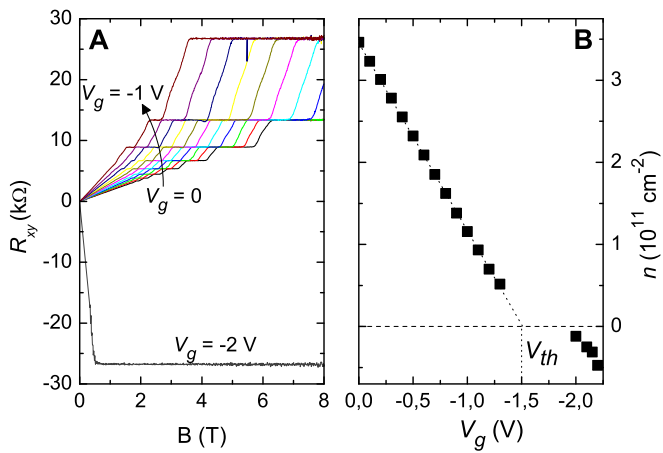}\\
 \caption{A: Hall resistance $R_{xy}$ as measured for various gate
voltages, indicating the transition from $n$- to $p$-conductance. B:
The gate-voltage dependent carrier density deduced from the Hall
measurements.}\label{Fignipdens}
\end{figure}

For the investigation of the QSH effect, samples with a low
intrinsic density, i.e., $n(V_g=0)<5\times10^{11}~{\rm cm}^{-2}$,
are studied. Transport measurements are done in a
$^3$He/$^4$He-dilution refrigerator with a base temperature
$T<30$~mK and a $^4$He cryostat ($T=1.4$~K) fitted with a vector
magnet system, which allows for magnetic fields up to 300~mT in
variable direction. When a negative gate voltage $V_g$ is applied to
the top gate electrode of the device, the usual decrease of the
electron density is observed. In Fig.~\ref{Fignipdens}~A,
measurements of the Hall resistance $R_{xy}$ are presented for a
Hall bar with length $L$ = $600~\mu$m and width $W$ = $200~\mu$m,
respectively. The decrease of the carrier density is reflected in an
increase of the Hall coefficient, when the gate voltage is lowered
from 0~V. For voltages down to -1.3~V, the density is decreased
linearly from $3.5\times10^{11}~{\rm cm}^{2}$ to
$0.5\times10^{11}~{\rm cm}^{2}$ (Fig.~\ref{Fignipdens}~B).
For even lower gate voltages, the sample becomes an insulator, with
the Fermi level in the band gap. Remarkably, when a large negative
voltage $V_g\leq-2$~V is applied, the sample becomes conducting
again. It can be inferred from the sign change of the Hall
coefficient that the device is now $p$-conducting. Obviously, the
Fermi level has now been pushed through the entire gap and into the
valence band.

The transition from $n$- to $p$-conductance is also reflected in the
behavior of the longitudinal resistance $R_{xx}$ as a function of
$V_g$ (cf. Fig.~\ref{FigQSHEnip}). For a decreasing electron density
the resistance rises until it reaches a maximum when the Fermi level
is in the gap. When the Fermi energy finally is pushed into the
valence band and the sample is $p$-conducting,
$R_{xx}$ decreases again by some orders of magnitude. A similar
transition from $n$- to $p$-conductance via an intermediate
high resistance regime has been observed for a variety of QW structures
with low intrinsic densities. The exact value of the maximum resistance, however,
depends critically on the nature of the band structure, and thus on
the actual QW width of the individual devices (cf. Section 7, below).

A similar signature in transport for a transition from $n$- to
$p$-conductance is possible and has been reported for the zero-gap
material graphene~\cite{Geim07}. However, since this material is a
semi-metal, no insulating regime is observed between $n$- and
$p$-conductance. Thus, the QSH effect is experimentally not
accessible in graphene even though it was explicitly predicted for
this material~\cite{kane2005A}. It was later understood
theoretically that a possible energy gap in graphene due to spin-orbit coupling
is far too small to be experimentally observable~\cite{yao2007,min2006}.

\begin{figure}[hbt]
\centering
 \includegraphics[width=0.8\linewidth]{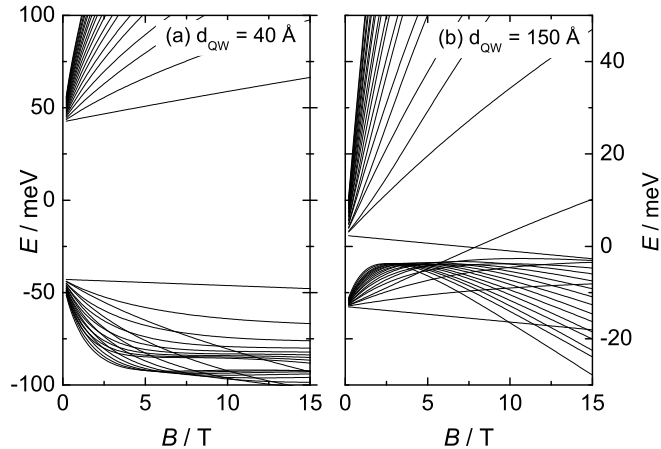}\\
 \caption{Landau level dispersion for (a) a 4.0 nm
 QW and (b) a 15.0 nm QW, respectively. The qualitative behavior is
 representative for samples with a normal and an inverted band
 structure, respectively.}
 \label{FigLLs}
\end{figure}

The peculiar band structure of inverted HgTe quantum wells gives rise to a
unique Landau level (LL) dispersion. For a normal band structure,
i.e., $d_{QW}<d_{c}$, all Landau levels are shifted to higher
energies for increasing magnetic field [Fig.~\ref{FigLLs}~(a)]. This
is the usual behavior and can be commonly observed for a variety of
materials.
When the band structure of the HgTe QW is inverted ($d_{QW}>d_{c}$),
the LL dispersion is markedly different [Fig.~\ref{FigLLs}~(b)].
Due to the mixing of the
electron-like and hole-like states, one of the states of the $H_1$
subband is a pure heavy-hole state ($M=-3/2$). Consequently, the
energy of the corresponding Landau level decreases with increasing magnetic
field. In adition, one of the valence band LLs has mainly electron
character and, thus, shifts to higher energies with magnetic field.
This leads to a crossing of these two peculiar LLs for a finite
magnetic field. The exact value of the crossing field ${\cal{B}}_{\rm
cross}$ depends on the QW width.
\begin{figure}[hbt]
\centering
 \includegraphics[width=0.8\linewidth]{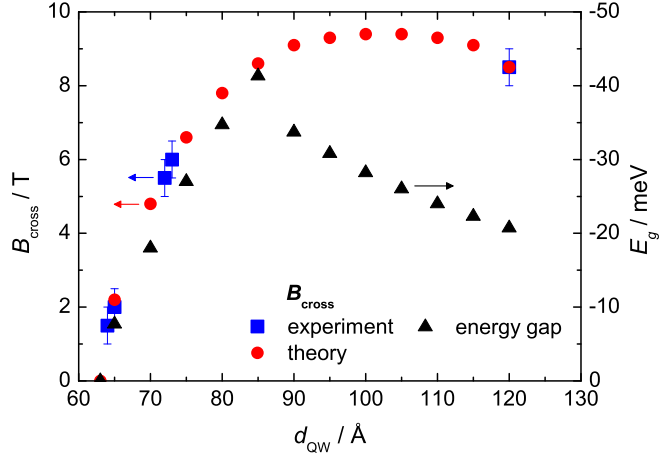}\\
 \caption{The ${\cal{B}}$-field value of the LL crossing determined
 experimentally (blue squares) and calculated theoretically (red
 circles). For the energy gap $E_g$ (black triangles), the negative
 values indicate the inverted band structure.}
 \label{FigBcross}
\end{figure}
Fig.~\ref{FigBcross} shows theoretically calculated
${\cal{B}}$-field values for the Landau level crossing for quantum
wells with an inverted band structure as red circles. The existence
of such a LL crossing is a clear indication for the occurrence of an
inverted band structure; for narrow QWs ($d_{QW}<d_{c}$) with a
normal band structure, there is no LL crossing. The figure also
plots the energy gap of the HgTe QWs, defined as $E_1 - H_1$, as
black triangles. In narrow wells ($d_{QW}<d_{c}$) with a normal band
structure, our definition yields a positive gap, whilst for inverted
wells we have an increasingly negative gap. For $d_{QW}\approx8.5$~
nm, the $E_1$ subband falls below the $H_2$ band (cf.
Fig.~\ref{FigEd}), and as a result the absolute value of the gap
decreases subsequently for larger QW widths.

\begin{figure}[p]
\centering
 \includegraphics[width=0.8\linewidth]{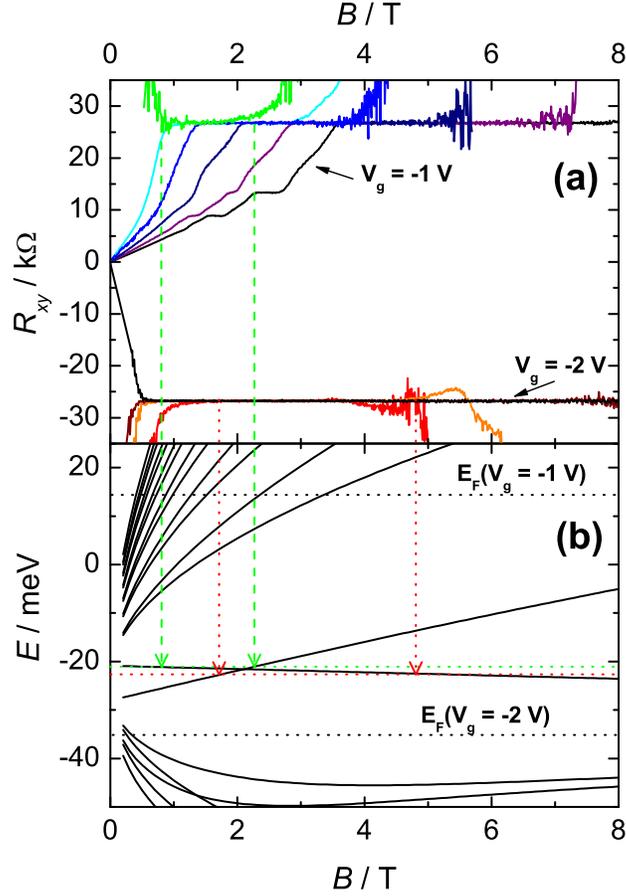}\\
 \caption{(a) Hall resistance, $R_{xy}$, of a $(L\times W) =
(600\times 200)$ $\mu$m$^2$
 QW structure with 6.5 nm well width for different carrier
concentrations obtained for
 different gate voltages $V_g$ in the range of -1~V to - 2~V. For
decreasing $V_g$,
 the $n$-type carrier concentration decreases and a transition to a
$p$-type conductor
 is observed, passing through an insulating regime  between -1.4~V and -1.9~V at
 ${\cal{B}}=0$~T.
(b) The Landau level fan chart of a 6.5~nm quantum well obtained
from an eight band {\bf k$\cdot$p} calculation. Black dashed lines
indicate the energetic position of the Fermi energy, $E_F$, for $V_g
= -1.0$~V and $-2.0$~V. Red and green dashed lines correspond to
position of the Fermi energies of the red and green Hall resistance
traces of (a). The crossing points of the Fermi level with the
respective Landau levels are marked by arrows of the same
color.}\label{FigexpBcross}
\end{figure}

The crossing of the conduction and valence band Landau levels can be
verified experimentally by quantum Hall experiments.
Fig.~\ref{FigexpBcross}~(a) shows quantum Hall data for an inverted
quantum well with a width of 6.5 nm. For gate voltages
$V_g\geq-1.0$~V and $V_g\leq-2.0$~V, the slope of the Hall signal
directly shows that the Fermi level is firmly in the conduction and
valence band, respectively. When the Fermi level is adjusted in the
gap, i.e., $-1.9$~V$<V_g<-1.4$~V, the Hall resistance shows
insulating behavior at zero field.

However, at higher magnetic fields the Hall resistance for these
gate voltages exhibits a re-entrance of the $n=1$ quantum Hall
plateau. This is a direct consequence of the crossing of the Fermi
level with the conduction-band derived LL for the green trace
($V_g=-1.4$~V) and with the valence band LL (e.g, for the red trace
where $V_g=-1.8$~V). As shown Fig.~\ref{FigexpBcross}~(b), the
experimental results are in good agreement with the theoretically
calculated LL dispersion. The crossing point of the Landau levels in
magnetic field, ${\cal{B}}_{\rm cross}$, can be determined
accurately by tuning the Fermi level through the energy gap. This
method has been used for various samples with an inverted band
structure; the results are included in Fig.~\ref{FigBcross} (blue
squares).

The observation of a re-entrant quantum Hall state is a clear
indication of the non-trivial insulating behavior, which is a
prerequisite for the existence of the QSH state. In contrast,
trivial insulating behavior is obtained for devices with
$d_{QW}<d_{c}$. For a normal band structure, the energy gap between
the lowest Landau levels of the conduction band and the valence
band, respectively, increases in magnetic field [cf.
Fig.~\ref{FigLLs}~(a)]. Thus, a sample remains insulating in
magnetic field, if the Fermi energy is located in the gap for
${\cal{B}}=0$.

\begin{figure}[p]
\centering
\includegraphics[scale=0.47]{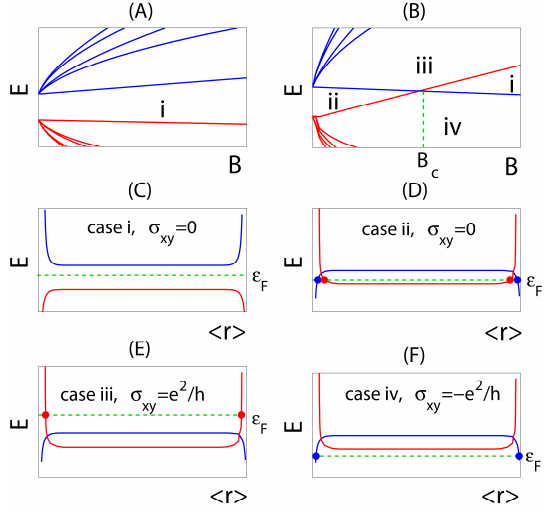}
\caption{(a) Landau level fan diagram for the $E1$ and $H1$ subbands
in the non-inverted regime. (b)Landau level fan diagram for the $E1$
and $H1$ subbands in the inverted regime. Notice that the zero-modes
from each set of subband Landau levels cross at a finite
${\cal{B}}$-field. Lower case roman numerals indicate separate
regions in energy and magnetic field. (c)-(f)Schematic energy
spectrum of the zero-modes and edge states for region i,ii,iii, and
iv respectively. (c)Region i produces a vanishing hall-conductance.
(d) Region ii produces a vanishing hall-conductance but the
Fermi-level passes through two branches of quantum Hall edge states
which are oppositely circulating and carry opposite current. (e)
Region iii produces a
 Hall-conductance of $+e^2/h,$ and the Fermi-level only crosses one of
the sets of edge states. (f) Region iv produces a
 Hall-conductance of $-e^2/h,$  and the Fermi-level only crosses one
of the sets of edge states.} \label{LLschem}
\end{figure}

The physics of this re-entrant quantum Hall state can basically be
understood within the simple $4$-band model. We will consider our
model given in Eq.~\ref{latticeH} expanded near ${\textbf{k}}=0$ and
in the presence of a uniform ${\cal{B}}$-field $(e{\cal{B}}>0)$ in
the $z$-direction. We choose the symmetric gauge with
${\textbf{A}}=\frac{{\cal{B}}}{2}(-y,x).$ We make the substitution
${\textbf{k}}\to ({\textbf{k}}+e{\textbf{A}})$ and define new
operators
\begin{eqnarray}
\pi_{+}&=&\hbar \left(k_{+}+\frac{ie{\cal{B}}}{2\hbar} z\right)\\
\pi_{-}&=&\hbar \left(k_{-}-\frac{ie{\cal{B}}}{2\hbar} z^{*}\right)
\end{eqnarray}\noindent where $k_{\pm}=(k_x\pm i k_y)$ and $z=x+i
y.$ These operators obey the commutation relations $\left[
\pi_{+},\pi_{-}\right]=\frac{-2\hbar^2}{\ell_{{\cal{B}}}^2}$ with
the magnetic length $\ell_{B}=(\hbar/e{\cal{B}})^{1/2}.$ Using these
commutation relations we can define raising and lowering operators
\begin{eqnarray}
a=\frac{\ell_{B}}{\hbar}\pi_{-}&,&\;\;\;a^{\dagger}=\frac{\ell_{B}}{\hbar}\pi_{+}\\
\left[a,a^{\dagger}\right]&=&1.
\end{eqnarray}\noindent Using these operators we rewrite our
Hamiltonian as
\begin{equation}
{\cal{H}}=\left(\begin{array}{cc} h_{+}(a^{\dagger},a)& 0\\
0&h_{-}(a^{\dagger},a)\end{array}\right)\end{equation}
\begin{eqnarray}
h_{+}(a^{\dagger},a)&=&(C-D)\frac{2}{\ell_{B}^2}(a^{\dagger}a^{\phantom{\dagger}}+\frac{1}{2})I_{2\times
2}
+(M-B)\frac{2}{\ell_{B}^2}(a^{\dagger}a^{\phantom{\dagger}}+\frac{1}{2})\sigma^z
+\frac{\sqrt{2}A}{\ell_B}(a^{\dagger}\sigma^{+}+a\sigma^{-})\nonumber\\
h_{-}(a^{\dagger},a)&=&(C-D)\frac{2}{\ell_{B}^2}(a^{\dagger}a^{\phantom{\dagger}}+\frac{1}{2})I_{2\times
2}+
(M-B)\frac{2}{\ell_{B}^2}(a^{\dagger}a^{\phantom{\dagger}}+\frac{1}{2})\sigma^z\nonumber\\
&-& \frac{\sqrt{2}A}{\ell_B}(a^{\dagger}\sigma^{-}+a\sigma^{+})
\label{bfieldH}\end{eqnarray}\noindent where
$\sigma^{\pm}=\frac{1}{2}(\sigma^x \pm i \sigma^y).$ It should be
noted in this context that the material-specific parameter $B$ of
the Dirac model, which plays a crucial role in describing the
re-entrant quantum Hall effect, is not related to the magnetic
field. The spectrum of this Hamiltonian can be solved since only a
finite number of harmonic oscillator Landau levels are coupled. We
find the energy spectrum\begin{equation} E_{\alpha}=-\omega_{0}^{D}
n -\alpha \frac{\omega_0^{B}}{2}\pm \sqrt{\frac{2A^2
n}{\ell_{B}^2}+(M-n\omega_0^{B}-\frac{\alpha}{2}\omega_{0}^{D})^2}
\end{equation}\noindent where $\omega_{0}^{B}=\frac{2}{\ell_{B}^2}B,$
$\omega_{0}^{D}=\frac{2}{\ell_{B}^2}D,\;$ $\alpha=\pm$ and
$n=0,1,2,\ldots$ This spectrum has ``zero modes" given by
\begin{eqnarray}
E^{0}_{+}&=&-M-\frac{1}{2}(\omega_{0}^{B}-\omega_{0}^D )\\
E^{0}_{-}&=&M+\frac{1}{2}(\omega_{0}^{B}+\omega_{0}^D )
\end{eqnarray}\noindent for $M/2B<0$ and
\begin{eqnarray}
E^{0}_{+}&=&M-\frac{1}{2}(\omega_{0}^{B}+\omega_{0}^D )\\
E^{0}_{-}&=&-M+\frac{1}{2}(\omega_{0}^{B}+\omega_{0}^D )
\end{eqnarray}\noindent for $M/2B>0.$ For $M/2B<0$ one can see
that the  set of zero-modes never cross. However, for $M/2B>0$,
i.e., in the inverted QSH regime, the zero modes do cross at a field
of ${\cal{B}}_{\rm cross}= \frac{\hbar}{e}\frac{M}{B+D}.$ For
typical parameters in the inverted
regime~\cite{Bernevig2006d,hughes2007} the critical field is just a
few Tesla. Since the Dirac mass $M$ reflects the energy gap, the
linear dependence of ${\cal{B}}_{\rm cross}$ on $M$ in the 4-band
Dirac model is in good agreement with the results obtained by the
$8\times8$ Hartree calculations shown in Fig.~\ref{FigBcross}. The
typical Landau level spectra for the non-inverted and inverted
regime, respectively, presented in Fig.~\ref{LLschem}~(a),(b) are
obtained by the simple 4-band model.

As illustrated in Fig.~\ref{LLschem} there are four different
regimes of interest. Two have quantized Hall conductances of $\pm
e^2/h$ and the other two have vanishing Hall conductances. Case
(i) is described by ${\cal{B}}>{\cal{B}}_c$, while the chemical
potential is placed between the zero modes. Here no edge states
exist and this yields  $\sigma_{xy}=0$. Identical behavior is
obtained for a trivial insulator independent of the
${\cal{B}}$-field strength, if the chemical potential is in the
energy gap. Case (ii) also shows $\sigma_{xy}=0$. However, there
should be counter-propagating edge states that carry opposite Hall
current, giving rise to the QSH effect. When the chemical
potential is slightly below the two zero modes as in case (iii),
only one of them will provide edge states. Thus, there is a single
state at each edge, resembling the well-known quantum Hall regime
with $\sigma_{xy}=e^2/h$. A similar result is obtained for case
(iv), where the chemical potential is just below the zero modes.
However, since the single edge state now originates from the other
zero mode, the Hall conductance changes sign, i.e.,
$\sigma_{xy}=-e^2/h$. If the chemical potential is fixed and
unequal to the energy of the LL crossing $E({\cal{B}}_{\rm
cross})$, an increase of the magnetic field gives rise to the
re-entrant quantum Hall state. While the Hall conductance vanishes
for case (i) and (ii), an intermediate regime with
$\sigma_{xy}=\pm e^2/h$ according to case (iii) and (iv),
respectively, will be obtained, while the sign of $\sigma_{xy}$
depends on the exact chemical potential.

\section{Experimental observation of the QSH effect at {${\cal{B}}=0$}}

\begin{figure}[hbt]
\centering
 \includegraphics[width=0.8\linewidth]{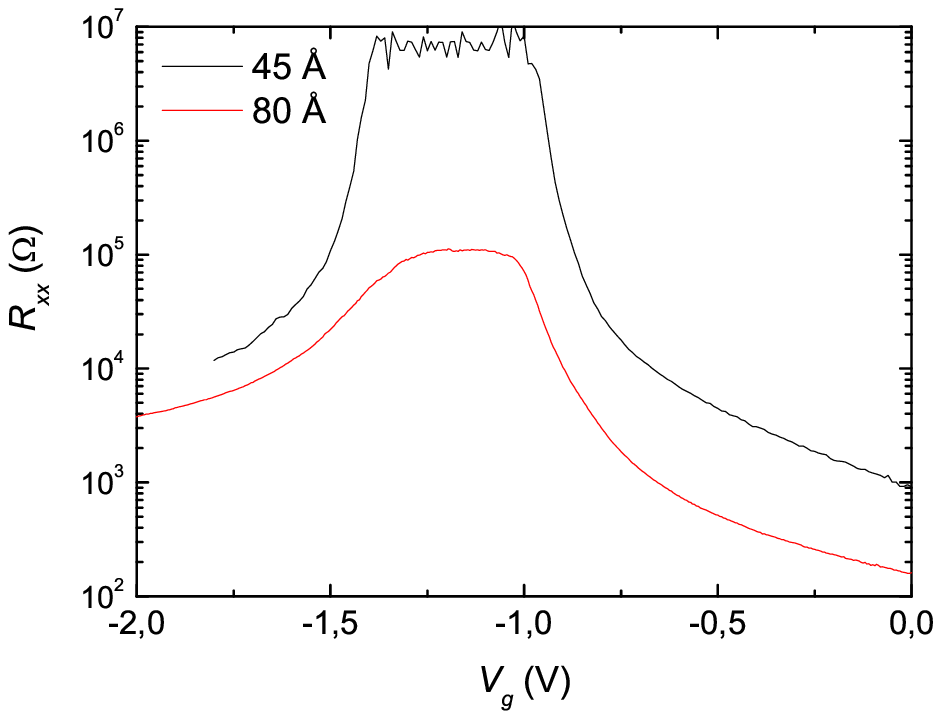}\\
 \caption{The longitudinal resistance of a 4.5 nm QW (a) and an
 8.0 nm QW (b) as a function of gate voltage, respectively.}
 \label{FigQSHEnip}
\end{figure}

 The existence of the QSH effect is revealed when
small Hall bars ($L\times W = 20.0~\mu{\rm m}\times13.3~\mu$m) are
studied. For QW devices with a normal band structure, the sample
shows trivial insulating behavior (Fig.~\ref{FigQSHEnip}). The
resistance of several M$\Omega$ for the insulating regime can be
attributed to the intrinsic noise level of the measurement setup, so
that the conductance basically vanishes.
For a device with an inverted band structure, however, the
resistance stays finite, not exceeding 100~k$\Omega$. This behavior
is reproduced for various Hall bars with a QW width in the range
from 4.5 nm to 12.0 nm: while devices with a normal band structure,
i.e., $d_{QW}<d_{c}\approx6.3$~nm, show trivial insulating behavior,
a finite conductance in the insulating regime is observed for
samples with an inverted band structure. We now proceed ro relate
this observed finite conductance to the predicted edge state
behavior. In the theoretical modelling of
Ref.~\citen{Bernevig2006d}, ideal contacts to the QSH edge states
were assumed, i.e., the potential of the edge states can be probed
without any influence due to the contacts. However, in reality the
leads are always $n$-type for the investigated samples and remain
such independently of the applied gate voltage. Unlike for a quantum
Hall system, the edge states in a QSH system propagate in both
directions at a given edge (cf. Fig.~\ref{FigHallbar}). Thus, the
edge states entering a contact carry a different potential due to
the different sources. When the edge states enter an $n$-type
contact, they necessarily equilibrate with the bulk states and
backscattering becomes possible. Hence, the voltage difference
between neighboring contacts does not vanish even though they are
connected by non-dissipative edge channels. In other words, a
voltage probe induces additional resistance, even when no voltage is
applied to it. Such a dissipation occurs due to the large size of
the $n$-type contact, which implies a huge number of channels the
incoming electron can scatter to. Consequently, an infinitesimal
dissipation in the contact can cause complete de-coherence between
the incoming and outgoing electrons. This analysis can be easily
confirmed by a Green's function calculation of the ballistic
conductance in the Keldysh formulism. In contrast, in a quantum Hall
system, a voltage probe does not induce additional dissipation,
because all the states are propagating to the same direction. The
de-coherence effect of a contact, although also it exists, does not
affect the current through it.

A simple Landauer-B{\"u}ttiker formalism~\cite{Buttiker86} can be
applied to determine the influence of the $n$-type contacts. The
current $I_i$ in a contact $i$ (cf. Fig.~\ref{FigHallbar}) can be
calculated from
$$I_i=(e/h)\sum_{j}T_{ij}(\mu_j-\mu_i).$$ Since the edge states are
non-interacting, the transmission coefficient $T_{ij}$ is unity for
neighboring contacts and zero otherwise. In the measurements, a
current $I$ is injected from contact 1 to contact 4, i.e.,
$I_1=-I_4=I$, while $I_i=0$ vanishes for all other contacts serving
as voltage probes. When total equilibration of all edge states in
each contact is assumed, a four-terminal resistance for neighboring
voltage probes of $(h/2e^2)$ is obtained. The two-terminal
resistance is determined by the number $n$ of the voltage probes
between the current source and drain. Between each pair of
neighboring contacts, a voltage $V_i=I\cdot(h/e^2)$ drops, adding up
to a total voltage drop of $(n+1)V_i$. Thus, the two-terminal
resistance is $(3h)/(2e^2)$ for a Hall bar geometry with two voltage
probes on each side as shown in Fig.~\ref{FigHallbar}.

However, the obtained finite resistance due to the $n$-type contacts
still does not give a quantitative explanation for the observed
resistance $R\approx100~{\rm k}\Omega$, which is significantly
higher than four terminal resistance $(h/2e^2) \approx 12.9~{\rm
k}\Omega$ one anticipates for the geometry used in the experiments.
The enhanced resistance in these samples with a length of
$L=20~\mu$m can be readily understood when inelastic scattering is
considered. While, as discussed above, the helical edge states are
robust against single-particle elastic backscattering, inelastic
mechanisms can cause backscattering.

For $n$-doped HgTe quantum wells, the typical mobility of the order
$10^5~{\rm cm}^2$/Vs implies an elastic mean free path of the order
of 1~$\mu$m~\cite{Daumer03}. Lower mobilities can be anticipated for
the QSH regime. The inelastic mean free path, which determines the
length scale of undisturbed transport by the QSH edge states, can be
estimated to be several times larger due to the suppression of
phonons and the reduced electron-electron interaction at low
temperatures. Thus, the inelastic scattering length is of the order
of a few microns.

For the observation of the QSH conductance, the sample dimensions
were reduced below the estimated inelastic mean free path. When
Hall bars with a length $L=1~\mu$m are studied, a four-terminal
resistance close to $h/(2e^2)$ is observed. The threshold voltage
$V_{th}$ is defined such that the QSH regime is in the vicinity of
$V_g=V_{th}$.
\begin{figure}[htb]
\centering
 \includegraphics[width=0.8\linewidth]{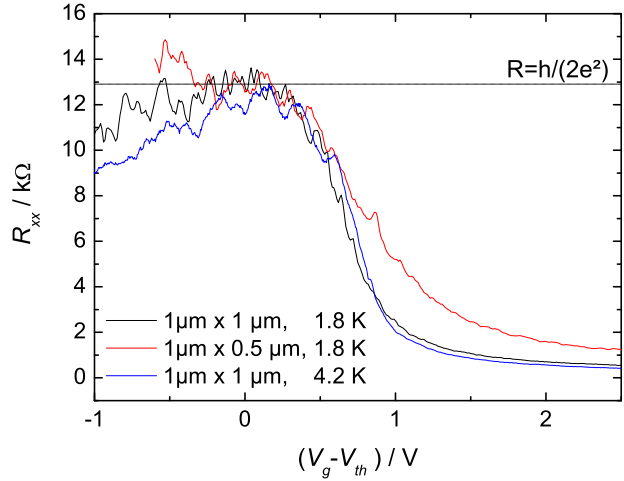}\\
 \caption{The longitudinal resistance is measured as a function of
 the gate voltage for two devices with $L=1\mu$m. The width $W$ is
 $1~\mu$m (black and blue) and $0.5~\mu$m (red), respectively. The black
 and red trace were obtained at 1.8~K, the blue one at 4.2~K.}
 \label{FigQSHEquant}
\end{figure}
The slight deviation of $R$ from the quantized value $h/(2e^2)$ can
be attributed to some residual scattering. This is an indication
that the length of the edge states still exceeds the inelastic mean
free path. However, a significant increase of the conductance is
obtained by reducing the sample size, which demonstrates clearly the
effect of scattering on the transport by QSH states.

The results presented in Fig.~\ref{FigQSHEquant} provide evidence
that transport in the QSH regime indeed occurs due to edge states.
The two devices with $W=1.0~\mu$m and $W=0.5~\mu$m, respectively,
were fabricated from the same QW structure. The resistance values of
the two devices differ significantly for the $n$-conducting regime,
where transport is determined by bulk properties. For the insulating
regime, however, both devices show the same resistance, even though
the width of the devices differs by a factor 2. For all three
measurements, no full transition to the $p$-conducting regime can be
observed. This is attributed to charging of the quantum well layer
during the e-beam process required to fabricate such small samples.
The fluctuations, which are observable in all traces, are
reproducible. Thus, they must have a physical reason, which is not
fully clear yet, and can not be caused, e.g., by electrical noise.
The pure QSH signal, however, is robust against an increase of
temperature as long as $k_BT$ is much smaller than the
two-dimensional bulk energy gap.

\section{Suppression of the quantum spin Hall effect by a magnetic field}

\begin{figure}[hbt]
\centering
 \includegraphics[width=0.8\linewidth]{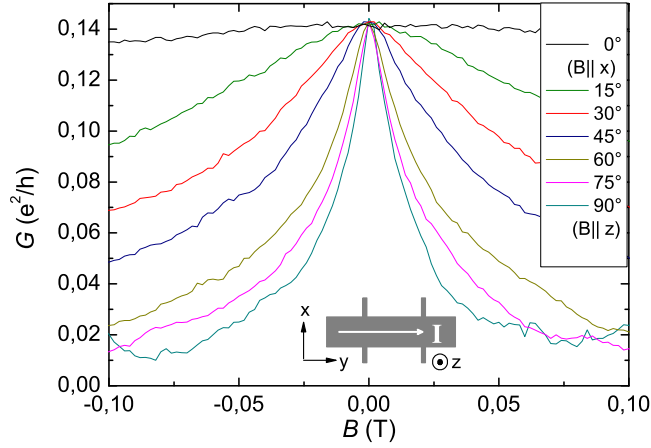}\\
 \caption{Four-terminal magnetoconductance, $G_{14,23}$,
 in the QSH regime as a function of tilt angle between
 the plane of the 2DEG and applied magnetic field for a
 $d = 7.3$ nm QW structure with dimensions
 $(L\times W = 20 \times 13.3)$ $\mu$m$^2$
 measured in a vector field cryostat at 1.4~K.}
 \label{FigQSHBanis}
\end{figure}

Further clear evidence for the QSH effect is obtained by
measurements in a finite magnetic field, which destroys the time
reversal symmetry of the QSH states. In order to observe this
effect, the dependence of the QSH conductance is studied in a
magnetic field in various directions with respect to the plane of
the 2DEG. For this purpose an inverted QW in Hall bar geometry
with $(L\times W) = (20.0\times13.3)~\mu{\rm m}^2$ was
investigated at 1.4~K in a vector magnet system. In a
perpendicular field configuration the QSH conductance decreases
rapidly when the magnetic field is applied. A cusp-like
conductance peak is observed with a full width-half maximum field
${\cal{B}}_{\rm FWHM}$ of 28~mT (Fig.~\ref{FigQSHBanis}), which is
even smaller at lower temperatures. At $T = 30$~mK a typical width
of ${\cal{B}}_{\rm FWHM}=10$~mT is observed. When the magnetic
field direction is rotated into the plane of the 2DEG the
conductance peak widens strongly, reaching a value of
${\cal{B}}_{\rm FWHM}\approx0.7$~T for the parallel configuration.

From the results shown in Fig.~\ref{FigQSHBanis} it is evident that
a perpendicular field has a much larger influence on the helical
liquid than an in-plane field.

Such a large anisotropy can be understood by a slightly modified
version of the four-band effective model we discussed in Sec. 3. To
 first order in perturbation theory, the gap induced by a perpendicular
and an in-plane field can be determined by the matrix elements of
the corresponding operator:
\begin{eqnarray}
E_{\rm gap\perp}&=&\left|\left\langle k=0,+\right|\left(\hat{\bf
z}\cdot\vec{\bf r}\times{\vec{\bf
j}}+ \mu_BM_z\right)\mathcal{B}_{\perp}\left|k=0,-\right\rangle\right|\nonumber\\
E_{\rm gap\parallel}&=&\left|\left\langle k=0,+\right|\mu_B\left(M_x
\mathcal{B}_x+M_y\mathcal{B}_y\right)\left|k=0,-\right\rangle
\right|
\end{eqnarray}
where $\vec{\bf r},~\vec{\bf j}$ are electron position and current
operators, respectively, and $\hat{\bf z}$ is the unit-vector
perpendicular to the 2DEG plane. The states
$\left|k=0,\pm\right\rangle$ are the two edge states propagating
toward opposite directions on the same boundary. The Zeeman
coupling matrix elements $M_{x,y,z}$ in the four-band effective
model can be determined by standard perturbation procedures from
the original Kane model, which has the form\cite{hughes2007}
\begin{eqnarray}
M_z&=&\left(\begin{array}{cccc}g_{E\perp}&&&\\&g_{H\perp}&&\\&&-g_{E\perp}&\\&&&-g_{H\perp}\end{array}\right)\nonumber\\
M_x&=&\left(\begin{array}{cccc}&&g_\parallel&\\&&&\\g_\parallel&&&\\&&&\end{array}\right),
M_y=\left(\begin{array}{cccc}&&-ig_\parallel&\\&&&\\ig_\parallel&&&\\&&&\end{array}\right).
\end{eqnarray}

Since the edge state wavefunctions of the effective model have been
obtained in Sec. 3, the gaps $E_{\rm gap\perp}$ and $E_{\rm
gap\parallel}$ can be calculated for given $g$-factors and Zeeman
couplings $S_{x,y,z}$. However, there is a subtlety that has to be
considered before carrying out this calculation. The effective model
(Eq.~\ref{contH}) consists of two decoupled blocks, which means the
coupling between $\left|E1,+\right\rangle,~\left|H1,+\right\rangle$
and $\left|E1,-\right\rangle,~\left|H1,-\right\rangle$ is ignored.
Thus for such an effective model, the current operator $\vec{\bf
j}=\nabla_{\bf k}H({\bf k})$ is also block diagonal. Since the
wavefunction of the edge state $\left|k,+(-)\right\rangle$ consists
only $\left|E1,+(-)\right\rangle$ and $\left|H1,+(-)\right\rangle$
components, the orbital magnetization operator $\hat{\bf
z}\cdot\vec{\bf r}\times{\vec{\bf j}}$ has a vanishing matrix
element between them, which means the orbital effect of the magnetic
field does not contribute to the gap $E_{\rm gap\perp}$. However,
this turns out to be an artifact of the simplified model, which
comes from the absence of bulk inversion asymmetry (BIA) terms. When
the breaking of bulk inversion symmetry is taken into account, an
additional term is induced in the four-band effective
model~\cite{hughes2007}:
\begin{eqnarray}
H_{\rm
BIA}=\left(\begin{array}{cccc}&&&-\Delta\\&&\Delta&\\&\Delta&&\\-\Delta&&&\end{array}\right)
\end{eqnarray}
which describes the mixing between $\left|E1,+(-)\right\rangle$
and $\left|H1,-(+)\right\rangle$. Such a term preserves
time-reversal symmetry, and does not close the bulk gap, and thus
does not affect the topological stability of the nontrivial
insulator phase. Consequently, all the theoretical discussions
based on the four-band model that we presented in the earlier part
of the paper are only slightly modified. Nevertheless, such a term
becomes important when we consider the effect of perpendicular
magnetic fields. Since the bulk-inversion symmetric terms have a
vanishing contribution to the edge state gap, the contribution
from BIA terms becomes the leading order. The BIA term $\Delta$
and all other parameters
$A,B,C,D,M,g_{E\perp},g_{H\perp},g_{\parallel}$ in the model can
be obtained from envelope function calculations of the quantum
well\cite{hughes2007}. For the quantum well thickness $d=7.0$~ nm,
these parameters are given as

\begin{center}
\begin{tabular}[h]{|c|c|c|c|c|c|c|c|}
\hline
$A({\rm eV\cdot \AA})$&$B({\rm eV\cdot \AA^2})$&$D({\rm eV})$&$M({\rm eV})$&$\Delta({\rm eV})$&$g_{E\perp}$&$g_{H\perp}$ &$g_{\parallel}$\\
\hline
$3.645$&$-68.6$&$-51.2$&$-0.010$&$0.0016$&$22.7$&$-1.21$&$-20.5$\\
\hline
\end{tabular}
\end{center}

By inputing these parameters into the tight-binding model
(Eq.~\ref{latticeH}) the edge state wavefunction can be obtained
numerically. For a magnetic field of $1{\rm T}$, the gap induced is
$E_{\rm gap\perp}=3.1$~meV if it is perpendicular to the 2DEG plane,
or $E_{\rm gap\parallel}=0.3$~meV if parallel. Since the
perpendicular and in-plane Zeeman coupling are of the same order, we
see that the main contribution of $E_{\rm gap\perp}$ comes from the
orbital effect of the magnetic field. Compared to the Zeeman
contribution, the orbital effect is equivalent to an ``effective
$g$-factor" of order $10^2$. Besides the numerical calculations, the
reason for such a large orbital effect can also be understood by a
rough order-of-magnitude estimation. Since the current operator
still doesn't mix $\left|E(H),+\right\rangle$ and
$\left|E(H),-\right\rangle$ even with BIA terms in the Hamiltonian,
the only contribution to $E_{\rm gap\perp}$ comes from the weight of
$\left|E(H),-\right\rangle$ components in the edge state
wavefunction $\left|k=0,+\right\rangle$, which is proportional to
$\Delta/|M|$ from first order perturbation theory. Thus we can
estimate the matrix element of the orbital magnetization operator as
$\left\langle k=0,+\right|\hat{\bf z}\cdot\vec{\bf r}\times\vec{\bf
j}\left|k=0,-\right\rangle\sim \left(\Delta/|M|\right)ev_F\xi$,
where we use $ev_F$ and the edge state width $\xi$ as the estimates
of operators $\vec{\bf j}$ and $\vec{\bf r}$, respectively. $v_F$
and $\xi$ of the edge states can be estimated by
$v_F=A/\hbar,~\xi\simeq A/|M|$. Thus $E_{\rm orb}\sim
\frac{e}{\hbar}\frac{\Delta A^2}{M^2}|\mathcal{B}_\perp|$. Compared
with the Bohr magnon we have the ``effective $g$-factor" as
\begin{eqnarray}
g_{\rm eff}\sim \frac{e}{\hbar}\frac{\Delta
A^2}{M^2}\frac1{\mu_B}=\frac{2m_ev_F^2}{M^2/\Delta}
\end{eqnarray}
Since $v_F=A/\hbar\simeq 5.5\times 10^5m/s$, we obtain $g_{\rm eff}$
is around $50$. Such an analysis also leads to the prediction that
the anisotropy between perpendicular and in-plane
magneto-conductance is further enhanced when the thickness goes
closer to the critical thickness $d_c$ from the inverted side, due
to the decrease of $|M|$.

\section{Conclusion}

We have reviewed our current theoretical understanding of the QSH
state, with a focus on the experimental realization in HgTe quantum
wells. We discussed the electronic structure of the HgTe quantum
wells in terms of a simple 4 band model which contains the essential
physics. We demonstrated the topological quantum phase transition as
the quantum well thickness $d_{QW}$ is varied and we explicitly
showed the analytic solution of the helical edge states on the
topologically non-trivial side of the phase transition. We discussed
the topological stability of the helical edge states using the
concrete example of the 4 band model for HgTe quantum wells and
presented the experimental realization of the QSH effect in HgTe QW
structures for various QW widths. These experiments clearly
demonstrated the edge channel character of the QSH effect and that
the inverted band structure, which occurs for HgTe QW with
$d_{QW}>6.3$~nm, is essential. Deviation from the quantized
conduction value for large samples and the effect of an applied
magnetic field could be related to back-scattering introduced by
time-reversal symmetry breaking processes.

The actual experimental realization of the QSH effect opens up the
opportunity for investigations of new theoretical
concepts~\cite{qi2007}, concerning fundamental aspects and
applications, utilizing the spin polarized properties of the helical
edge channels.

We wish to thank B.A. Bernevig, X. Dai, Z. Fang and C.J. Wu for
insightful discussions, S. Wiedmann for assistance with the
experiments, A. Roth, C. Br{\"u}ne, C.R. Becker and V. Hock for
sample preparation, and C. Kumpf for calibrating the well widths
of the HgTe samples. This work is supported by the DFG (SFB 410),
by the German-Israeli Foundation for Scientific Research and
Development (Grant No.881/05), by NSF through the grants
DMR-0342832, and by the US Department of Energy, Office of Basic
Energy Sciences under contract DE-AC03-76SF00515, and Focus Center
Research Program (FCRP) Center on Functional Engineered
Nanoarchitectonics (FENA).

\bibliographystyle{unsrt}
\bibliography{hgte,review}

\begin{thebibliography}{10}

\bibitem{prinz1998}
G.~Prinz.
\newblock {\em Science}, 282:1660, 1998.

\bibitem{wolf2001}
\textrm{S. A. Wolf} \emph{et. al.}
\newblock {\em Science}, 294:1488, 2001.

\bibitem{murakami2003}
S.~Murakami, N.~Nagaosa, and \textrm{S.C. Zhang}.
\newblock 301:1348, 2003.

\bibitem{sinova2004}
\textrm{J. Sinova} \emph{et. al.}
\newblock {\em Phys. Rev. Lett.}, 92:126603, 2004.

\bibitem{kato2004}
\textrm{Y. Kato} \emph{et. al.}
\newblock {\em Nature}, 427:50, 2004.

\bibitem{wunderlich2005}
J.~Wunderlich, B.~Kaestner, J.~Sinova, and T.~Jungwirth.
\newblock {\em Phys. Rev. Lett.}, 94:47204, 2005.

\bibitem{dyakonov1971}
\textrm{M.I. Dyakonov} and \textrm{V.I. Perel}.
\newblock {\em Sov. Phys. JETP}, 13:467, 1971.

\bibitem{hirsch1999}
J.E. Hirsch.
\newblock {\em Phys. Rev. Lett.}, 83:1834, 1999.

\bibitem{thouless1982}
\textrm{D.J. Thouless}, M.~Kohmoto, \textrm{M.P. Nightingale}, and \textrm{M.
  den Nijs}.
\newblock {\em Phys. Rev. Lett.}, 49:405, 1982.

\bibitem{murakami2004a}
S.~Murakami, N.~Nagaosa, and \textrm{S.C. Zhang}.
\newblock {\em Phys. Rev. Lett.}, 93:156804, 2004.

\bibitem{kane2005A}
\textrm{C. L. Kane} and \textrm{E. J. Mele}.
\newblock {\em Phys. Rev. Lett.}, 95:226801, 2005.

\bibitem{bernevig2006A}
\textrm{B.A. Bernevig} and \textrm{S.C. Zhang}.
\newblock {\em Phys. Rev. Lett.}, 96:106802, 2006.

\bibitem{wu2006}
\textrm{C. Wu}, \textrm{B.A. Bernevig}, and \textrm{S.C. Zhang}.
\newblock {\em Phys. Rev. Lett.}, 96:106401, 2006.

\bibitem{xu2006}
\textrm{C. Xu} and \textrm{J. Moore}.
\newblock {\em Phys. Rev. B}, 73:045322, 2006.

\bibitem{kane2005B}
\textrm{C. L. Kane} and \textrm{E. J. Mele}.
\newblock {\em Phys. Rev. Lett.}, 95:146802, 2005.

\bibitem{yao2007}
Yugui Yao, Fei Ye, Xiao-Liang Qi, Shou-Cheng Zhang, and Zhong Fang.
\newblock {\em Phys. Rev. B}, 75:041401(R), 2007.

\bibitem{min2006}
H.~Min, J.~Hill, N.~Sinitsyn, B.~Sahu, L.~Kleinman, and A.~MacDonald.
\newblock {\em Phys. Rev. B}, 74:165310, 2006.

\bibitem{Bernevig2006d}
\textrm{B. A. Bernevig}, \textrm{T. L. Hughes}, and \textrm{S.C. Zhang}.
\newblock {\em Science}, 314:1757, 2006.

\bibitem{Novik05}
E.~G. Novik, A.~Pfeuffer-Jeschke, T.~Jungwirth, V.~Latussek, C.~R. Becker,
  G.~Landwehr, H.~Buhmann, and L.~W. Molenkamp.
\newblock {\em Phys. Rev. B}, 72:035321, 2005.

\bibitem{Pfeuffer}
A.~Pfeuffer-Jeschke.
\newblock PhD thesis, Universit{\"a}t W{\"u}rzburg, 2000.

\bibitem{kane1957}
\textrm{E.O. Kane}.
\newblock {\em J. Phys. Chem. Solids}, 1:249, 1957.

\bibitem{redlich1984}
\textrm{A. N. Redlich}.
\newblock {\em Phys. Rev. D}, 29:2366, 1984.

\bibitem{volovik2003}
G.~Volovik.
\newblock {\em The Universe in a Helium Droplet}.
\newblock Oxford Publications, Oxford, 2003.

\bibitem{qi2005}
\textrm{X.L. Qi}, \textrm{Y.S. Wu}, and \textrm{S.C. Zhang}.
\newblock {\em Phys. Rev. B}, 74:085308, 2006.

\bibitem{creutz1994}
\textrm{M. Creutz} and \textrm{I. Horvath}.
\newblock {\em Phys. Rev. D}, 50:2297, 1994.

\bibitem{creutz2001}
\textrm{M. Creutz}.
\newblock {\em Rev. Mod. Phys.}, 73:119, 2001.

\bibitem{wen1990}
X.~G. Wen.
\newblock {\em Phys. Rev. B}, 41:12838, 1990.

\bibitem{NIELSEN1981}
H.~B. Nielsen and M.~Ninomiya.
\newblock {\em Nucl. Phys. B}, 185:20, 1981.

\bibitem{NIELSEN1981A}
H.~B. Nielsen and M.~Ninomiya.
\newblock {\em Nucl. Phys. B}, 193:173, 1981.

\bibitem{fu2006}
Liang Fu and C.~L. Kane.
\newblock {\em Phys. Rev. B}, 74(19):195312, 2006.

\bibitem{fu2007a}
Liang Fu and C.~L. Kane.
\newblock {\em Phys. Rev. B}, 76(4):045302, 2007.

\bibitem{fu2007b}
Liang Fu, C.~L. Kane, and E.~J. Mele.
\newblock {\em Phys. Rev. Lett.}, 98(10):106803, 2007.

\bibitem{moore2007}
\textrm{J. E. Moore} and \textrm{L. Balents}.
\newblock {\em Phys. Rev. B}, 75:121306, 2007.

\bibitem{roy2006a}
Rahul Roy.
\newblock arxiv: cond-mat/0604211.

\bibitem{roy2006b}
Rahul Roy.
\newblock arxiv: cond-mat/0607531.

\bibitem{roy2006c}
Rahul Roy.
\newblock arxiv: cond-mat/0608064.

\bibitem{niu1985}
Qian Niu, D.~J. Thouless, and Yong-Shi Wu.
\newblock {\em Phys. Rev. B}, 31:3372, 1985.

\bibitem{essin2007}
Andrew~M. Essin and J.~E. Moore.
\newblock {\em Physical Review B}, 76:165307, 2007.

\bibitem{lee2007}
Sung-Sik Lee and Shinsei Ryu.
\newblock arxiv:0708.1639.

\bibitem{Becker07}
C.~R. Becker, C.~Br{\"u}ne, M.~Sch{\"a}fer, A.~Roth, H.~Buhmann, and L.~W.
  Molenkamp.
\newblock {\em phys. stat. sol. (c)}, 4:3382, 2007.

\bibitem{Daumer03}
V.~Daumer, I.~Golombek, M.~Gbordzoe, E.~G. Novik, V.~Hock, C.~R. Becker,
  H.~Buhmann, and L.~W. Molenkamp.
\newblock {\em Appl. Phys. Lett.}, 83:1376, 2003.

\bibitem{Zhang01}
X.~C. Zhang, A.~Pfeuffer-Jeschke, K.~Ortner, V.~Hock, H.~Buhmann, C.~R. Becker,
  and G.~Landwehr.
\newblock {\em Phys. Rev. B}, 63:245305, 2001.

\bibitem{Hinz06}
J.~Hinz, H.~Buhmann, M.~Sch{\"a}fer, V.~Hock, C~R. Becker, and L.~W. Molenkamp.
\newblock {\em Semicond. Sci. Technol.}, (21):501, 2006.

\bibitem{Geim07}
A.~K. Geim and K.~S. Novoselov.
\newblock {\em Nature Materials}, 6:183, 2007.
\newblock and references therein.

\bibitem{hughes2007}
Taylor~L. Hughes, Chao-Xing Liu, Xiao-Liang Qi, and Shou-Cheng Zhang.
\newblock in preparation.

\bibitem{Buttiker86}
M.~B{\"u}ttiker.
\newblock {\em Phys. Rev. Lett.}, 57:1761, 1986.

\bibitem{qi2007}
Xiao-Liang Qi, Taylor~L. Hughes, and Shou-Cheng Zhang.
\newblock arxiv:0710.0730.

\end{thebibliography}
\end{document}